\documentclass[prb,nofootinbib,twocolumn,superscriptaddress]{revtex4} 

\usepackage{graphicx}
\usepackage{dcolumn}
\usepackage{bm}
\usepackage{threeparttable}
\usepackage{times}
\usepackage{mathptmx}
\usepackage{lscape}
\usepackage{natbib}
\usepackage{amsmath}
\usepackage{amssymb}
\usepackage{braket}
\usepackage{comment}
\usepackage{color}

\def\degree{\kern-.2em\r{}\kern-.3em}

\begin{document}


\title{ Where Nonlinearity in Thermodynamic Average Comes from? \\ Configurational Geometry Revisited   }

\author{Koretaka Yuge}
\affiliation{
Department of Materials Science and Engineering,  Kyoto University, Sakyo, Kyoto 606-8501, Japan\\
}%

\author{Shouno Ohta}
\affiliation{
Department of Materials Science and Engineering,  Kyoto University, Sakyo, Kyoto 606-8501, Japan\\
}%

\author{Ryogo Miyake}
\affiliation{
Department of Materials Science and Engineering,  Kyoto University, Sakyo, Kyoto 606-8501, Japan\\
}%

\begin{abstract}
{  For classical discrete system under constant composition, we theoretically examine origin of nonlinearity in thermodynamic (so-called canonical) average w.r.t. many-body interactions, in terms of geometrical information in configuratin space. We clarify that nonlinearity essentially comes from deviation in configurational density of states (CDOS) \textit{before} applying many-body interactions to the system, from multidimensional gaussian distribution. The present finding strongly suggest the significance to investigate how the deviation in CDOS bridges bidirectional stability relationships between equilibrium structure and potential energy in thermodynamic average. 
  }
\end{abstract}


\maketitle

\section{Introduction}
For classical discrete system under constant composition typically refered to substitutional alloys, expectation value of structure along chosen coordination $p$ under given coordination $\left\{ Q_{1}, \cdots, Q_{f} \right\}$ is expressed as
\begin{eqnarray}
\label{eq:can}
\Braket{ Q_{p}}_{Z} = Z^{-1} \sum_{i} Q_{p}^{\left( i \right)} \exp \left( -\beta U^{\left( i \right)} \right),
\end{eqnarray}
where $\Braket{\quad}_{Z}$ denotes canonical average, $Z=\sum_{i}\exp\left(-\beta U_{i}\right)$ partition function, $\beta$ inverse temperature, and summation is taken over all possible microscopic states on configuration space. 
One of the most natural selection for the coordination is employing complete orthonormal basis (COB), which is typically constructed by generalized Ising model (GIM),\cite{ce} which is considered in the present case. When we employ such COB, potential energy $U$ for microscopic structure $k$ is exactly given by
\begin{eqnarray}
\label{eq:u}
U^{\left( k \right)} = \sum_{j} \Braket{U|Q_{j}} Q_{j}^{\left( k \right)},
\end{eqnarray}
where $\Braket{\quad|\quad}$ denotes inner product, i.e., trace over possible states on configuration space. 
When we introduce two $f$-dimensional vectors of $\mathbf{Q}_{Z}=\left(\Braket{ Q_{1}}_{Z},\cdots, \Braket{ Q_{f}}_{Z}\right)$ and $\mathbf{U}=\left(U_{1},\cdots,U_{f}\right)$ ($U_{b}=\Braket{U|Q_{b}}$), it is clear from 
Eqs.~\eqref{eq:can} and~\eqref{eq:u} that generally, $\mathbf{Q}_{Z}$ is not a linear function of $\mathbf{U}$, i.e., thermodynamic (here, canonical) average is a nonlinear map w.r.t. potential energy. 
Due to the nonlinear character, it is typically difficult to exactly determine temperature dependence of $\Braket{Q}_{Z}$ for given potential energy: Thus, various approaches have been developed including Metropolis algorism, entropic sampling and Wang-Landau sampling for efficient exploration of important microscopic states to determine equilibrium properties.\cite{mc1,mc2,wl} 
Very recently, we quantitatively formulate bidirectional stability (BS) character of thermodynamic average between equilibrium structure and potential energy surface in terms of their hypervolume correspondence, where the nonlinearity plays essential role to break the BS character.\cite{bd} 
Despite such significance, origin of the nonlinearity in terms of configurational geometry, i.e., geometric information in configuration space without requiring any thermodynamic information such as temperature or energy, has not been addressed so far. 
We here show that the nonlinearity comes from any deviation in configurational density of states \textit{before} applying many-body interaction to the system, from ideal Gaussian distribution. The details are shown below.

\section{Derivation and Application}
\subsection{Derivation for condition of thermodynamic average as a linear map}
Hereinafter for simpliticy (without loss of generality), structure (value of $Q$s) are measured from that at center of gravity of configurational density of states (CDOS), given by $\left\{\Braket{Q_{1}},\cdots,\Braket{Q_{f}}\right\}$, where $\Braket{\quad}$ denotes taking linear average for all possible microscopic states. 
If the thermodynamic average is a linear map, it should be given by
\begin{eqnarray}
\label{eq:lin}
\mathbf{Q}_{Z} = \beta \cdot \Lambda \cdot \mathbf{U}, 
\end{eqnarray}
where $\Lambda$ is temperature-independent $f\times f$ matrix, and $\beta$ in r.h.s. should be required from dimensional analysis between $Q$ and $U$. 
On the other hand, Mclaughlin expantion of $\Braket{Q_{p}}_{Z}$ at $\beta=0$ up to $g$-th order is given by
\begin{widetext}
\begin{eqnarray}
\label{eq:mac}
\mathrm{M}\left[\Braket{Q_{p}}_{Z} \right]= \left. \frac{\partial \ln Z}{\partial\left(-u_{p}\right)} \right|_{\beta=0}  + \sum_{n=1}^{g}  \left[ \frac{1}{n!}\sum_{k_{1},\cdots,k_{n}=1}^{g} \left(-u_{k_{1}}\right)\cdots \left(-u_{k_{n}} \right) \left\{ \left.  \frac{\partial}{\partial\left(-u_{k_{1}}\right)}\cdots\frac{\partial}{\partial\left(-u_{k_{n}}\right)} \frac{\partial}{\partial\left(-u_{p}\right)} \ln Z \right|_{\beta=0}  \right\}  \right],
\end{eqnarray}
\end{widetext}
where $u_{p}=\beta \cdot \Braket{U|Q_{p}}$.
For the Mclaughlin expantion, when we assume that partial differentiation of $\ln Z$ by $\left\{ -u_{k_{\left( 1,1 \right)}},\cdots,-u_{k_{\left( 1,m_{1} \right)}},\cdots,-u_{k_{\left( n,m_{n} \right)}} \right\}$ ($u_{k_{\left( q,r \right)}}$ takes one of $\left\{ u_{1},\cdots, u_{f} \right\}$ where duplicate selection is allowed) is given by the following products for canonical average
\begin{eqnarray}
\label{eq:partial}
\Braket{ Q_{k_{\left( 1,1 \right)}}\cdots Q_{k_{\left( 1,m1 \right)}} }_{Z}\cdots \Braket{ Q_{k_{\left( n,1 \right)}} \cdots Q_{k_{\left( n, m_{n} \right)}}}_{Z},
\end{eqnarray}
partial differentiation of Eq.~\eqref{eq:partial} by $-u_{k}$ becomes
\begin{eqnarray}
\label{eq:qm}
&&\Braket{ Q_{k_{\left( 1,1 \right)}}\cdots Q_{k_{\left( 1,m1 \right)}}\cdot Q_{k} }_{Z}\cdots \Braket{ Q_{k_{\left( n,1 \right)}} \cdots Q_{k_{\left( n, m_{n} \right)}}}_{Z} \nonumber \\
&+& \cdots \nonumber \\
&+& \Braket{ Q_{k_{\left( 1,1 \right)}}\cdots Q_{k_{\left( 1,m1 \right)}} }_{Z}\cdots \Braket{ Q_{k_{\left( n,1 \right)}} \cdots Q_{k_{\left( n, m_{n}\right)}}\cdot Q_{k} }_{Z} \nonumber \\
&-& \Braket{ Q_{k_{\left( 1,1 \right)}}\cdots Q_{k_{\left( 1,m1 \right)}} }_{Z}\cdots \Braket{ Q_{k_{\left( n,1 \right)}} \cdots Q_{k_{\left( n, m_{n} \right)}}}_{Z} \Braket{Q_{k}}_{Z}. \nonumber \\
\quad
\end{eqnarray}
When we apply partial differentiation of $\ln Z$ by $-u_{k}$, we get
\begin{eqnarray}
\label{eq:zu}
\frac{\partial \ln Z}{\partial \left( -u_{k} \right) } = \Braket{Q_{k}}_{Z}.
\end{eqnarray}
From Eqs.~\eqref{eq:partial}-\eqref{eq:zu}, we can  see that $\ln Z$ can be partial differentiated by any times for $-u_{k}$, which leads to that we can take $g\to\infty$ in Eq.~\eqref{eq:mac}.
We here focus on the condition where 
\begin{eqnarray}
\label{eq:inf}
\lim_{g\to\infty} \mathrm{M}\left[ \Braket{Q_{p}}_{Z} \right] = \Braket{Q_{p}}_{Z}
\end{eqnarray}
is satisfied. 
Then, by comparing Eqs.~\eqref{eq:lin} and~\eqref{eq:mac}, we can immediately see that when thermodynamic average is a linear map, coefficients for $\beta^{r}$ ($r\ge 2$) in Eq.~\eqref{eq:mac} should be all zero. 
From Eqs.~\eqref{eq:partial}-\eqref{eq:zu}, it is clear that partial derivatives of $\ln Z$ in Eq.~\eqref{eq:mac} by combination of $\left\{ -u_{k_{1}},\cdots,-u_{k_{m}} \right\}$ should always results in summation of multiple terms consisting of products for canonical average, where each term contains every \textit{single} $Q_{k_{1}},\cdots, Q_{k_{m}}$. 
For instance, terms for $\beta$ is given by 
\begin{eqnarray}
-\sum_{k_{1}} u_{k_{1}} \Braket{Q_{p}Q_{k_{1}}}, 
\end{eqnarray}
and those for $\beta^{2}$, $\beta^{3}$ and $\beta^{4}$ in Eq.~\eqref{eq:mac} respectively becomes
\begin{widetext}
\begin{eqnarray}
\label{eq:beta}
&&\frac{1}{2}\sum_{k_{1},k_{2}} u_{k_{1}}u_{k_{2}}\Braket{Q_{p}Q_{k_{1}}Q_{k_{2}}}, \quad -\frac{1}{6} \sum_{k_{1},k_{2},k_{3}}u_{k_{1}}u_{k_{2}}u_{k_{3}}\left( \Braket{Q_{p}Q_{k_{1}}Q_{k_{2}}Q_{k_{3}}} - \sum_{\left\{ j \right\}} \Braket{Q_{j_{1}} Q_{j_{2}} } \Braket{Q_{j_{3}}Q_{j_{4}}}    \right), \nonumber \\
&&\frac{1}{24} \sum_{k_{1},k_{2},k_{3},k_{4}} u_{k_{1}}u_{k_{2}}u_{k_{3}}u_{k_{4}} \left( \Braket{Q_{p}Q_{k_{1}}Q_{k_{2}}Q_{k_{3}}Q_{k_{4}}}  - \sum_{\left\{ j \right\}} \Braket{Q_{j_{1}}Q_{j_{2}}Q_{j_{3}}}\Braket{Q_{j_{4}}Q_{j_{5}}} \right),
\end{eqnarray}
\end{widetext}
where summation $\left\{ j \right\}$ takes all possible combination including index $p$ and the rest from $\left\{ k \right\}$ where duplicate selection is not allowed. 
Generally, terms for $\beta^{r}$ corresponds to polynomial for $\left\{ u_{k_{1}}\cdots u_{k_{r}} \right\}$, where its coefficient is given by linear combination of $\left( r+1 \right)$-th order multivariate moment $\Braket{Q_{p}Q_{k_{1}}\cdots Q_{k_{r}}}$ and products of lower-order moments, as seen is Eq.~\eqref{eq:beta}. 
Therefore, if thermodynamic average is a linear map w.r.t. any given set of $u_{k}$, $t$-th ($t\ge 3$) multivariate moments should either take  zero or be given by linear combination of lower-order moments. For instance with $t=4$, we obtain
\begin{eqnarray}
\label{eq:mom}
\forall p, \quad \Braket{Q_{p}Q_{i}Q_{k}Q_{l}} = \sum_{\left\{ j \right\}} \Braket{Q_{j_{1}}Q_{j_{2}}}\Braket{Q_{j_{3}}Q_{j_{4}}}.
\end{eqnarray}
We have recently clarified that when CDOS is exactly given by multidimensional Gaussian distribution, thermodynamic average exactly becomes linear map by the form of Eq.~\eqref{eq:lin}.\cite{em2} This directly means that all $t$-th ($t\ge 3$) multivariate moments of Gaussian satisfy terms for $\beta^{t-1}$ taking zero, seen in Eq.~\eqref{eq:mom}. 
From above discussions, when provided CDOS leads to thermodynamic average as linear map, relationships between $t$-th ($t\ge 3$) order multivariate moments and second-order moments of $\left\{ \Braket{Q_{i}Q_{k}}|i,k=1,\cdots,f \right\}$ (i.e., covariance matrix of $\Gamma_{ik} = \Braket{Q_{i}Q_{k}}$) should be exactly same as the relationships for multidimensional Gaussian. 
Therefore, if a certain Gaussian can have the same covariance matrix $\Gamma$, the provided CDOS should be nothing but Gaussian. Since covariance matrix is by definition real symmetric, any $\Gamma$ can always be diagonalized with appropriate orthogonal matrix $\mathbf{P}$ of
\begin{eqnarray}
\mathbf{C} = \mathbf{P}^{-1}\Gamma\mathbf{P},
\end{eqnarray}
with diagonal elements of $\left\{ c_{1},\cdots,c_{f} \right\}$.
Let us consider whether or not $\left\{d_{i}\right\}$ includes zero value. 
When we choose lineary independent basis under constant composition, it always corresponds to including emply $Q_{\textrm{E}}$ (i.e., independent of composition and of configuration) and multisite correlations: This leads to a set of basis function of $\left\{ Q_{\textrm{E}},Q_{1},\cdots, Q_{f} \right\}$. 
If $\Gamma$ for $\left\{ Q_{1},\cdots, Q_{f} \right\}$ has zero eigenvalue, one of the basis after orthogonal transformation (with fixing $Q_{\textrm{E}}$) is independent of configuration, i.e., it takes constant value.  
In this case, such a basis should clearly be linear dependent with $Q_{\textrm{E}}$, where such coodrination should be omitted for basis under given constant composition. Therefore, for any practical CDOS under such basis, $\Gamma$ should always be positive definite. 
Under this coordination, multidimensional Gaussian can be simply constructed by product of a set of single-variate Gaussian with variance of $d_{k}$. 
These directly means that for classical discrete system with complete basis, we can always construct multidimensional Gaussian with any given $\Gamma$ obtained from practical CDOS. 
Therefore, under the condition of Eq.~\eqref{eq:inf}, CDOS providing thermodynamic average as a linear map is restricted to multidimensional Gaussian with covariance matrix of $\Gamma$.



\subsection{Application to practical systems}
Very recently, we quantitatively formulate bidirectional stability relationships $B$ in thermodynamic average between equilibrium structure and potential energy, based on its nonlinearity:
\begin{eqnarray}
B= \log \left| 1+ \mathrm{div} \mathbf{D} + \sum_{\mathrm{F}}J_{\mathrm{F}}\left[ \frac{\partial \mathbf{D}}{ \partial Q} \right] + J\left[ \frac{\partial \mathbf{D}}{ \partial Q} \right] \right|,
\end{eqnarray}
where $\mathbf{D}$ is called ``anharmonicity in structural degree of freedom'', which is a vector field on configuration space defined as
\begin{eqnarray}
\mathbf{D}\left( Q \right) &=& \left( D_{1}\left( Q \right),\cdots,D_{f}\left( Q \right) \right) \nonumber \\
D_{i}\left( Q \right) &=&  \left\{\left(\phi_{\textrm{th}}\left( \beta \right)\circ \left( -\beta\cdot\Lambda \right)^{-1} \right)\cdot Q - Q\right\}_{i}.
\end{eqnarray}
Here, $\phi_{\textrm{th}}$ denotes canonical average, and $\left\{ \quad \right\}_{i}$ represents $i$-th component of $\mathbb{R}^{f}$ vector. Since $\Lambda$ is a invertible linear map and image of composite map $\phi_{\textrm{th}}\left( \beta \right)\circ \left( -\beta\cdot\Lambda \right)^{-1} $ is exactly independent of energy and of temperature, $\mathbf{D}$ is a measure of nonlinearity for $\phi_{\textrm{th}}$ depending only on configurational geometry \textit{before} applying many-body interaction to the system, i.e., can be known a priori without any thermodynamic information. 
Therefore, from above discussions, it is fundanemtally important to investigate how the deviation in CDOS from gaussian connects with vector field $\mathbf{D}$.

\begin{figure}
\begin{center}
\includegraphics[width=0.9\linewidth]{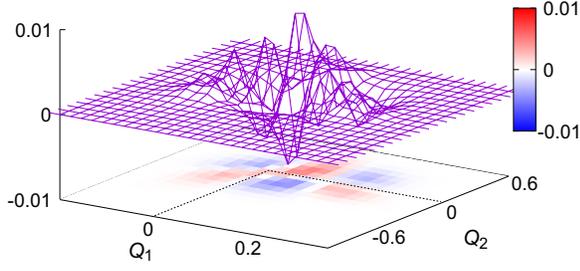}
\caption{Deviation in CDOS on fcc equiatomic binary system from Gaussian distribution.}
\label{fig:dcdos}
\end{center}
\end{figure}
In order to qualitatively address this point, we first rewrite canonical average of Eq.~\eqref{eq:can} by using CDOS of $n\left( Q_{1},\cdots,Q_{f} \right)$, namely
\begin{eqnarray}
\label{eq:ncan}
\Braket{Q_{p}}_{Z} = \frac{\displaystyle{\int} n\left( Q_{1},\cdots,Q_{f} \right)Q_{p}\exp{\left(-\beta\sum_{j=1}^{f} \Braket{U|Q_{j} }Q_{j} \right)} d\mathbf{Q} }{\displaystyle{\int}  n\left( Q_{1},\cdots,Q_{f} \right)\exp{\left(-\beta\sum_{j=1}^{f} \Braket{U|Q_{j} }Q_{j} \right)} d\mathbf{Q} }. \nonumber \\
\quad
\end{eqnarray}
Therefore, $\Braket{Q_{p}}_{Z}$ can be interpret as a \textit{functional} of CDOS, $n\left( Q_{1},\cdots,Q_{f} \right)$.
Then, corresponding functional derivative is immediately given by
\begin{eqnarray}
\label{eq:fd}
\frac{\delta\Braket{Q_{p}}_{Z}}{\delta n } = \frac{\left( Q_{p} - \Braket{Q_{p}}_{Z}^{\left( n \right)} \right) \cdot \exp{\left(-\beta\sum_{j=1}^{f} \Braket{U|Q_{j} }Q_{j} \right)  }  }{ Z^{\left( n \right)} }, \nonumber \\
\quad
\end{eqnarray}
where $Z^{\left( n \right)}$ and $\Braket{Q_{p}}_{Z}^{\left( n \right)}$ respectively denotes partition function and canonical average of $Q_{p}$ for CDOS of $n\left( Q_{1},\cdots,Q_{f} \right)$. 
To apply the functional derivative of Eq.~\eqref{eq:fd} to investigating the behavior of vector field $\mathbf{D}$, we first numerically obtain \textit{exact} CDOS on fcc equiatomic binary system with 32-atom ($2\times 2\times 2$ expansion of conventional 4-atom unit cell) along 1NN and 2NN pair correlation, $Q_{1}$ and $Q_{2}$, by constructing all possible ${}_{32}\mathrm{C}_{16}$ atomic configurations.
\begin{figure}
\begin{center}
\includegraphics[width=0.63\linewidth]{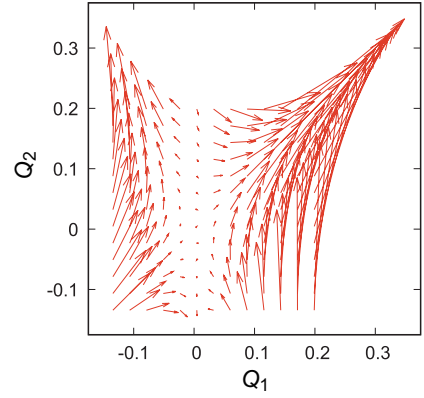}
\caption{Simulated vector field $\mathbf{D}$.}
\label{fig:D}
\end{center}
\end{figure}
\begin{figure}
\begin{center}
\includegraphics[width=0.74\linewidth]{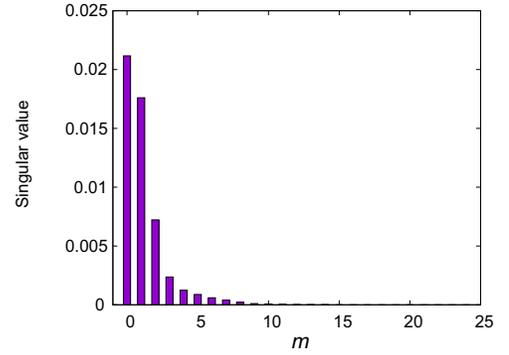}
\caption{Singular values for deviation in CDOS from Gaussian.}
\label{fig:svd}
\end{center}
\end{figure}
Figure~\ref{fig:dcdos} shows the resultant CDOS measured from corresponding Gaussian with the same covariance matrix. 
We can clearly see the character around center of gravity: (i) CDOS at 1st and 3rd quadrunts shows positive, and that at 2nd and 4th quadrunts shows negative deviation, and (ii) CDOS at 1st and 4th quadrunts have larger absolute deviation than at 2nd and 3rd quadrunts. 
Figure~\ref{fig:D} shows vector field $\mathbf{D}$ near center of gravity, using the information about the obtained exact CDOS. 

We now would like to know how the behavior of $\mathbf{D}$ in Fig.~\ref{fig:D} and deviation in CDOS of Fig.~\ref{fig:dcdos} are bridged, based on functional derivative of Eq.~\eqref{eq:fd}.
To achieve this, we apply singular value decomposition (SVD) to the deviation in CDOS, whose singular values are shown in Fig.~\ref{fig:svd}: We can see that the deviation in CDOS can be appproximated as low-rank discrete function.

\begin{figure}[h]
\begin{center}
\includegraphics[width=0.9\linewidth]{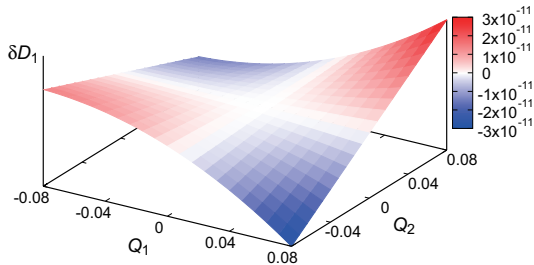}
\caption{ Contribution from changes in CDOS of $f\left( Q_{1}, Q2 \right)$  to the behavior of $D_{1}$.}
\label{fig:dD}
\end{center}
\end{figure}

We therefore simplify the deviation in CDOS from Gaussian to capture the above features as 
\begin{eqnarray}
h\left( Q_{1}, Q_{2} \right) \simeq h\left( Q_{1} \right)h\left( Q_{2} \right),
\end{eqnarray}
where
\begin{eqnarray}
h\left( Q_{1} \right) &=& c_{1}\delta\left( Q_{1} - Q_{k1} \right) - c'_{1}\delta\left( Q_{1} + Q_{k1} \right) \nonumber \\
h\left( Q_{2} \right) &=& c_{2}\delta\left( Q_{2} - Q_{k2} \right) - c_{2}\delta\left( Q_{2} + Q_{k2} \right).
\end{eqnarray}
Here, $Q_{k1}$, $Q_{k2}$, $c_{1}$, $c'_{1}$ and $c_{2}$ are positive constants with $c_{1}>c'_{1}$, and $\delta\left( Q \right)$ is a Dirac delta function.
We perform inner product between the functional derivative and the deviation in CDOS to see the contribution from the simplyfied changes $h\left( Q_{1},Q_{2} \right)$ contributes to $D_{1}$, namely
\begin{widetext}
\begin{eqnarray}
\label{eq:fin}
\delta D_{1}^{\left( h \right)}\left( Q_{1},Q_{2} \right) &&\equiv \Braket{\frac{\delta\Braket{Q_{p}}_{Z}}{\delta n },h\left( Q_{1},Q_{2} \right) }_{g} \nonumber \\
&&= \frac{-2c_{2}\cdot \sinh\left( -\Lambda_{22}^{-1}Q_{2}Q_{k2} \right)}{Z^{\left( g \right)}\left( Q_{1},Q_{2} \right) }\left[ c_{1} \left( Q_{k1} - Q_{1} \right) \exp\left( \Lambda_{11}^{-1}Q_{1} \right)  +  c'_{1} \left( Q_{k1} + Q_{1} \right) \exp\left( -\Lambda_{11}^{-1}Q_{1} \right)  \right], 
\end{eqnarray}
\end{widetext}
where 
\begin{eqnarray}
Z^{\left( g \right)} = \frac{1}{\sqrt{\Lambda_{11}\Lambda_{22}} }\exp\left( \frac{ \Lambda_{11}^{-1}Q_{1}^{2} + \Lambda_{22}^{-1} Q_{2}^{2} }{2 }   \right).
\end{eqnarray}
Here, $\Braket{\quad}_{g}$ means taking inner product with gaussian CDOS, and substitute the relationships in ASDF of $Q_{1} \simeq -\beta \Lambda_{11} \Braket{U|Q_{1}}$ and $Q_{2} \simeq -\beta \Lambda_{22} \Braket{U|Q_{2}}$ (approximation comes from neglecting off-diagonal elements of $\Lambda$ just for simplicity of Eq.~\eqref{eq:fin}, which does not provide significant effect in the present study). 
Figure~\ref{fig:dD} shows the resultant behavior of $\delta D_{1}^{\left( h \right)}$ with $c_{1}=c_{2}=2c'_{1}$ and $Q_{k1}=Q_{k2}=10e-4$, which mimics the condition that slight changes in CDOS having the similar feature of Fig.~\ref{fig:dcdos} appears near the center of gravity. 
We can clearly see that behavior of $\delta D_{1}^{\left( h \right)}$ can qualitatively capture that of $D_{1}$ for the sign of each quadrunts. 
Since landscape of $\delta D_{1}^{\left( h \right)}$ depend on $f\left( Q_{1},Q_{2} \right)$, further investigation of vector field $\mathbf{D}$ would require decomposition of $f\left( Q_{1},Q_{2} \right)$ into low-rank functions, e.g., based on SVD. 

\section{Conclusions}
We examine the origin of nonlinearity in thermodynamic average for classical discrete systems under constant composition, in terms of configurational geometry. We clarify that the nonlinearity comes from any deviation in configurational density of states \textit{before} applying many-body interaction to the system, from ideal Gaussian distribution with the same covariance matrix as practical CDOS. The present results strongly indicate profound relationships between deviation in CDOS from Gaussian and bidirectional stability character in thermodynamic average.

\section{Acknowledgement}
This work was supported by Grant-in-Aids for Scientific Research on Innovative Areas on High Entropy Alloys through the grant number JP18H05453 and a Grant-in-Aid for Scientific Research (16K06704) from the MEXT of Japan, Research Grant from Hitachi Metals$\cdot$Materials Science Foundation, and Advanced Low Carbon Technology Research and Development Program of the Japan Science and Technology Agency (JST).

\end{document}